\documentclass[11pt]{article}
\usepackage{fullpage,latexsym,amsmath,amssymb}

\newcommand{\newfontobj}[2]{
  \newcommand{#1}[1]{\expandafter\def\csname##1\endcsname{{#2 ##1}}}}

\newfontobj{\class}{\bf}

\renewcommand{\P}{{\bf P}}

\class{NP}
\class{PH}
\class{NQP}
\class{FP}
\class{SPP}
\class{BPP}
\class{BQP}
\class{EQP}
\class{AWPP}
\class{AAWPP}
\class{GUM}
\class{GAM}
\class{GUAM}
\class{PUM}
\class{PAM}
\class{PUAM}
\class{UP}
\class{AM}
\class{MA}
\class{PP}
\class{NC}
\class{AC}
\class{ACC}
\class{QNC}
\class{EQNC}
\class{NQNC}
\class{BQNC}
\class{QACC}
\class{QAC}
\class{NQAC}
\class{NQACC}
\class{TC}
\class{QTC}

\newcommand{\SAT}{{\rm SAT}}

\newcommand{\ints}{{\mathbb Z}}

\newcommand{\map}[3]{{#1}:{#2}\rightarrow{#3}}

\newcommand{\ket}[1]{|{#1}\rangle}
\newcommand{\tuple}[1]{\langle{#1}\rangle}
\newcommand{\bigtuple}[1]{\left\langle{#1}\right\rangle}

\newcommand{\prob}[1]{{\sc #1}}
\newcommand{\OC}{\prob{Orbit Coset}}

\newcommand{\HSh}{\prob{Hidden Shift}}
\newcommand{\HC}{\prob{Hidden Coset}}
\newcommand{\GHSh}{\prob{Generalized Hidden Shift}}
\newcommand{\ghsh}{\prob{GHSh}}

\newcommand{\Gint}{\prob{Group Intersection}}

\newcommand{\GI}{\prob{Graph Isomorphism}}
\newcommand{\SLV}{\prob{Shortest Lattice Vector}}
\newcommand{\HS}{\prob{Hidden Subgroup}}

\newcommand{\etalchar}[1]{$^{#1}$}

\newtheorem{definition}{Definition}[section]
\newtheorem{proposition}[definition]{Proposition}
\newtheorem{lemma}[definition]{Lemma}
\newtheorem{claim}[definition]{Claim}
\newtheorem{theorem}[definition]{Theorem}
\newtheorem{corollary}[definition]{Corollary}

\newenvironment{proof}{\noindent{\bf Proof.}}{\hfill$\Box$ \bigskip}
\newenvironment{proofof}[1]{\noindent{\bf Proof of
{#1}.}}{\bigskip}

\title{The central nature of the Hidden Subgroup problem}

\author{Stephen Fenner\thanks{Department of Computer Science \& Engineering, Columbia, SC 29208 USA\@.  {\tt fenner@cse.sc.edu}.  Partially supported by NSF grant CCF-0515269.}\\
University of South Carolina
\and
Yong Zhang\thanks{Department of Mathematical Sciences, 1200 Park Road, Harrisonburg, VA 22802-2462 USA\@.  {\tt yong.zhang@emu.edu}.  Partially supported by an EMU Summer Research Grant, 2006.}\\
Eastern Mennonite University}

\date{September 18, 2006}

\begin{document}
\maketitle

\begin{abstract}
We show that several problems that figure prominently in quantum computing, including {\HC}, {\HSh}, and
{\OC}, are equivalent or reducible to {\HS} for a large variety of
groups.  We also
show that, over permutation groups, the decision version and search
version of {\HS} are polynomial-time
equivalent.  For {\HS} over dihedral
groups, such an equivalence can be obtained if the order of the group is smooth.  Finally, we give nonadaptive program checkers for {\HS}
and its decision version.
\end{abstract}

\noindent{\bf Topic Classification}: Computational Complexity, Quantum Computing.

\section{Introduction}

The {\HS} problem
generalizes many interesting problems that have
efficient quantum algorithms but whose known classical
algorithms are inefficient.  While we can solve {\HS} over abelian 
groups satisfactorily on quantum computers, the nonabelian case
is more challenging.  Until now only limited success has been
reported.
For a recent survey on the progress of solving
nonabelian {\HS}, see Lomont
\cite{lomont04:_hidden_subgr_probl_review_and_open_probl}.
People are particularly interested in solving {\HS} over
two families of nonabelian groups---permutation groups and dihedral groups---since
solving them will immediately give solutions to the
{\GI} problem \cite{jozsa00:_quant} and the {\SLV}
problem \cite{regev04:_quant}, repectively. 

To explore more fully the power of quantum computers, researchers have also
introduced and studied several related problems.  Van 
Dam, Hallgren, and Ip \cite{dam03:_quant} introduced the {\HSh} problem
and gave efficient quantum algorithms for some instances.  Their results provide evidence that quantum computers can help to 
recover shift structure as well as subgroup structure.  They
also introduced the {\HC} problem to generalize {\HSh} and {\HS}.  Recently, Childs and van Dam \cite{childs05:_quant} introduced the {\GHSh} problem, which extends {\HSh} from a different angle.  They gave efficient quantum algorithms for {\GHSh} over cyclic groups where the number of functions is large (see Definition~\ref{def:group-problems} and the subsequent discussion).  In an attempt to
attack {\HS} using a divide-and-conquer approach over subgroup chains, Friedl et al.\ \cite{friedl03:_hidden} introduced the {\OC} problem, which they claimed to be an even more general problem including {\HS} and {\HSh}\footnote{They actually called it the Hidden Translation problem.} as special instances.  They called {\OC} a \emph{quantum} generalization of {\HS} and {\HSh}, since the definition of {\OC} involves \emph{quantum functions}.

In Section~\ref{reduction}, we show that all these related problems are 
equivalent or reducible to {\HS}.  In particular,
\begin{enumerate}
\item
{\HC} is polynomial-time equivalent to {\HS},
\item {\OC} is equivalent to {\HS} if we allow
functions in the latter to be quantum functions, and
\item
{\HSh} and {\GHSh} reduce to instances of {\HS} over a family of wreath product groups.
\end{enumerate}
Some special cases of these results are already known.  It is well-known that {\HSh} over the cyclic group $\ints_n$ is equivalent to {\HS} over the dihedral group $D_n = \ints_n \rtimes \ints_2$ (see \cite{childs05:_quant} for example), and this fact easily generalizes to any abelian group.  Our results apply to general groups,  however, including nonabelian groups where a nontrivial semidirect product with $\ints_2$ may not exist.  Regarding the relationship between {\HSh} and {\GHSh}, Childs and van Dam observed that it is trivial to reduce any instance of {\GHSh} to {\HSh} over the same group (and thence to dihedral {\HS} in the case of abelian groups) in polynomial time \cite{childs05:_quant}.  They left open the question, however, of whether any versions of {\GHSh} with more than two functions are \emph{equivalent} to any versions of {\HS}.  We make progress towards answering this question in the affirmative.  We give a direct ``embedding'' reduction from {\GHSh} to {\HS} such that the original input instances of {\GHSh} can be recovered efficiently from their images under the reduction.  Our reduction runs in polynomial time provided the number of functions of the input instance is relatively small.

There are a few results in the literature about the complexity of {\HS}\@.  It is well-known that {\HS}
over abelian groups is solvable in quantum polynomial time with bounded error \cite{kitaev95:_abelian_stabilizer,mosca99:_thesis}.
Ettinger, Hoyer, and
Knill \cite{ettinger04} showed that {\HS} (over arbitrary finite groups) has polynomial quantum query complexity.
Arvind and Kurur \cite{AK:GIinSPP} 
showed that {\HS} over permutation groups is in the class
$\FP^{\SPP}$ and is thus low for the counting complexity class
$\PP$\@. In Section~\ref{search} we 
study the relationship between the decision and search
versions of {\HS}, denoted {\HS}$_D$ and {\HS}$_S$, respectively. 
It is well known that $\NP$-complete sets such as {\SAT} are
self-reducible, which implies that the decision
and search versions of $\NP$-complete problems are polynomial-time equivalent.  We show
this is also the case for {\HS} and {\HSh} over permutation
groups.  Kempe and Shalev have recently given evidence that {\HS}$_D$ over permutation groups is a difficult problem \cite{kempe_shalev05:_permutation_hsp}.  They showed that under general conditions, various forms of the Quantum Fourier Sampling method are of no help (over classical exhaustive search) in solving {\HS}$_D$ over permutation groups.  Our results yield evidence of a different sort that this problem is difficult---namely, it is just as hard as the search version.

For {\HS} over dihedral groups, our results are more modest.  We show the search-decision equivalence for dihedral groups of
smooth order, i.e., where the largest prime dividing the order of the group is small.

Combining our results in Sections~\ref{reduction} and \ref{search},
we obtain nonadaptive program checkers for {\HS} and {\HS}$_D$
over permutation groups.  We give the details in Section~\ref{programchecker}.

\section{Preliminaries}

\subsection{Group Theory}

Background on general group theory and quantum computation can be found in 
textbooks such as \cite{scott87:_group_theor} and \cite{NC:quantumbook}. 

The wreath product of groups plays an important role in several proofs
in this paper.  We only need to define a special case of the wreath product.

\begin{definition}\rm
For any finite group $G$, 
the \emph{wreath product} $G \wr \ints_n$ of $G$ and 
$\ints_n= \{ 0, 1, \ldots, n-1 \}$ is the set $\{(g_1, g_2, \ldots, g_n, \tau) \mid
g_1, g_2,\ldots,g_n \in
G, \; \tau \in \ints_n \}$ equipped with the group operation $\circ$ such that
\[ (g_1, g_2, \ldots,g_n,\tau) \circ (g_1', g_2',\ldots,g_n', \tau') = (g_{\tau'(1)}g_1',
g_{\tau'(2)}g_2',\ldots, g_{\tau'(n)}g_n',\tau\tau' ). \]
We abuse notation here by identifying $\tau$ and
$\tau'$ with cyclic permutations over the set $\{1,\ldots,n\}$ sending $x$ to $x+\tau \bmod n$ and to $x + \tau' \bmod n$, respectively, and identifying $0$ with $n$.
\end{definition}

If $Z$ is a set, then $S_Z$ is the \emph{symmetric group} of permutations of $Z$. We define the composition order to be from
left to right, i.e., for $g_1, g_2 \in S_Z$, $g_1g_2$ is the permutation obtained 
by applying $g_1$ first and then $g_2$.  For $n\ge 1$,
we abbreviate $S_{\{1,2,\ldots,n\}}$ by $S_n$.  Subgroups of $S_n$ are the
\emph{permutation groups} of degree $n$.  For a permutation group $G\le S_n$ and an element 
$i\in \{1,\ldots,n\}$, let $G^{(i)}$ denote the pointwise stabilizer
subgroup of $G$ that fixes the set $\{1,\ldots,i\}$ pointwise.  The
chain of the stabilizer subgroups
of $G$ is $\{id\}=G^{(n)}\leq G^{(n-1)}\leq \cdots \leq G^{(1)} \leq G^{(0)} = G$.
Let $C_i$ be a complete set of right coset representatives of $G^{(i)}$ in 
$G^{(i-1)}$, $1\leq i\leq n$. Then the cardinality of $C_i$ is at most
$n-i$ and $\cup_{i=1}^{n} C_i$ forms a \emph{strong generator set}
for $G$ \cite{sims70:_comput}. 
Any element $g\in G$ can be written uniquely as $g=g_ng_{n-1}\cdots g_1$ with
$g_i \in C_i$. Furst, Hopcroft, and Luks \cite{furst80:_polyn} showed that given
any generator set for $G$, a strong generator set can be computed in polynomial time.
For $X \subseteq Z$ and $G \leq S_Z$, we use $G_X$ to denote the
subgroup of $G$ that stablizes $X$ setwise.  It is evident that $G_X$ is the direct sum of $S_X$ and $S_{Z \setminus X}$.  We are particularly
interested in the case when $G$ is $S_n$.  In this case, a generating
set for $G_X$ can be easily computed.

Let $G$ be a finite group. Let $\Gamma$ be a set of mutually
orthogonal quantum states. Let $\map{\alpha}{G\times\Gamma}{\Gamma}$
be a group action of $G$ on $\Gamma$, i.e., for every $x\in G$ the
function $\map{\alpha_x}{\Gamma}{\Gamma}$ mapping $\ket{\phi}$ to $\ket{\alpha (x,\ket{\phi})}$
is a permutation over $\Gamma$, and the map $h$ from $G$ to the symmetric group over $\Gamma$ defined by $h(x)=\alpha_x$ is a homomorphism.  We use the notation $\ket{x\cdot \phi}$ instead of $\ket{\alpha
(x,\ket{\phi})}$, when $\alpha$ is clear from the context.  We let
$G(\ket{\phi})$ denote the orbit of $\ket{\phi}$ with respect to $\alpha$, i.e., the set $\{\ket{x\cdot\phi}: x\in G\}$, and we
let $G_{\ket{\phi}}$ denote the stabilizer subgroup of $\ket{\phi}$ in
$G$, i.e., $\{x\in G: \ket{x\cdot \phi} = \ket{\phi} \}$.  Given any
positive integer $t$, let $\alpha^t$ denote the group action of $G$ on
$\Gamma^t=\{\ket{\phi}^{\otimes t} : \ket{\phi}\in \Gamma\}$ defined
by $\alpha^t (x,\ket{\phi}^{\otimes t}) = \ket{x\cdot\phi}^{\otimes
t}$. We need $\alpha^t$ because the input superpositions cannot be
cloned in general.

\begin{definition}\label{def:group-problems}\rm
Let $G$ be a finite group.
\begin{enumerate}
\item\label{item:HS-def} Given a generating set for $G$ and a function $f$ that maps $G$ to some finite set $S$ such that the values of $f$ are constant on a subgroup $H$ of $G$ and distinct on each left (right) coset of $H$, the \emph{{\HS} problem} is to find a generating set for $H$.  The \emph{decision version} of {\HS}, denoted as {\HS}$_D$, is to determine whether $H$ is trivial.  The \emph{search version}, denoted as {\HS}$_S$, is to find a nontrivial element of $H$ if there is one.
\item\label{item:GHSh-def} Given a generating set for $G$ and $n$
injective functions $f_1, f_2, \ldots, f_n$ defined on $G$, with the
promise that there is a (necessarily unique) ``shift'' $u \in G$ such
that for all $g \in G$, $f_1(g) = f_2(ug)$, $f_2(g)=f_3(ug)$, \ldots,
$f_{n-1}(g)=f_n(ug)$, the \emph{{\GHSh} problem} \cite{childs05:_quant} is to find $u$.  We sometimes denote this problem as \emph{$(n,G)$-\ghsh} for short.  If $n=2$, this problem is called the \emph{{\HSh} problem}.  The functions $f_1, \ldots, f_n$ are given uniformly via a single function $F$ such that $f_i(g) = F(i,g)$ for all $g\in G$ and $1\le i \le n$.
\item\label{item:HC-def} Given a generating set for $G$ and two
functions $f_1$ and $f_2$ defined on $G$ such that for some shift
$u\in G$, $f_1(g)=f_2(gu)$ for all $g$ in $G$, the \emph{{\HC}
problem} \cite{dam03:_quant} is to find the set of all such shifts $u$.  This set is a coset $Hu$ of a subgroup $H$ of $G$, and we can represent it by giving generators for $H$ together with one of the $u$.
\item\label{item:OC-def} Given a generating set for $G$ and two
quantum states $\ket{\phi_0}, \ket{\phi_1}\in \Gamma$, the \emph{{\OC} problem} \cite{friedl03:_hidden} is to
either reject the input if $G(\ket{\phi_0})\cap G(\ket{\phi_1})=\emptyset$, or else output both a $u \in G$ such that $\ket{u \cdot \phi_1} = \ket{\phi_0}$ and also a generating set for $G_{\ket{\phi_1}}$.
\end{enumerate}
\end{definition}

Van Dam, Hallgren, and Ip give efficient quantum algorithms for
various instances of {\HC} using Fourier sampling \cite{dam03:_quant}.
Childs and van Dam give a polynomial-time quantum algorithm for
$(M,\ints_N)$-{\ghsh} when $M \ge N^{\epsilon}$ for any fixed
$\epsilon > 0$ \cite{childs05:_quant}.  Friedl, et al.\
\cite{friedl03:_hidden} give polynomial-time quantum algorithms for
(among others) $(2,\ints_p^n)$-{\ghsh} where $p$ is a fixed prime, and
more generally for $(2,G)$-{\ghsh} if $G$ is ``smoothly solvable,'' a
class of groups that includes solvable groups of bounded exponent and
bounded derived series length.  The latter results come via algorithms
for {\OC}.

\subsection{Program checkers}

Let $\pi$ be a computational decision or search problem. Let $x$ be an
input to $\pi$ and $\pi(x)$ be the output of $\pi$. Let $P$ be a
deterministic program (supposedly) for $\pi$ that halts on all inputs.  
We are interested in whether $P$ has any bug, i.e., whether there is
some $x$ such that $P(x)\neq \pi(x)$. A efficient \emph{program checker} $C$
for $P$ is a probabilistic expected-polynomial-time oracle Turing
machine that uses $P$ as an oracle and takes $x$ and a positive
integer $k$ (presented in unary) 
as inputs. The running time of $C$ does not include the time it takes
for the oracle $P$ to do its computations.
$C$ will output CORRECT with probability $\geq 1-1/2^k$ if
$P$ is correct on all inputs (no bugs), and output BUGGY with probability $\geq 1-1/2^k$ if
$P(x)\neq \pi(x)$. This probability is over the sample space of all
finite sequences of coin flips $C$ could have tossed. However, if $P$ has bugs
but $P(x)=\pi(x)$, we allow $C$ to 
behave arbitrarily. If $C$ only queries the oracle nonadaptively, then
we say $C$ is a \emph{nonadaptive checker}.
See Blum and Kannan \cite{blum95:_desig} for more details.

\section{Several Reductions} \label{reduction}

The {\HC} problem is to find the set of all shifts of the two functions $f_1$ and $f_2$ defined on the group $G$.  If $Hu$ is the coset of all shifts, then $f_1$ is constant on $H$ (see \cite{dam03:_quant} Lemma 6.1).  If we let $f_1$ and $f_2$ be the same function chosen appropriately, we get {\HS} as a special case.  On the other hand, if $f_1$ and $f_2$ are injective functions, this is {\HSh}.
 
\begin{theorem}\label{HC}
{\HC} is polynomial-time equivalent to {\HS}.
\end{theorem}

\begin{proof}
Let $G$ and $f_1,f_2$ be the input of {\HC}.  Let the set of shifts be
$Hu$, where $H$ is a subgroup of $G$ and $u$ is a coset representative.  Define a function $f$ with domain $G \wr \ints_2$ as follows: for any $(g_1, g_2, \tau) \in G \wr \ints_2$, 
\[ f(g_1, g_2, \tau) = \left\{ \begin{array}{cl}
    (f_1(g_1), f_2(g_2)) & \mbox{if $\tau=0$,}\\
    (f_2(g_2), f_1(g_1)) & \mbox{if $\tau=1$.}
\end{array} \right.  \]
The values of $f$ are constant on the
set $K = (H \,\times\, u^{-1}Hu \,\times \,\{0\}) \cup (u^{-1}H\, \times \,Hu \,\times\, \{1\})$, which is a subgroup of $G \wr \ints_2$. Furthermore, the values of $f$ are distinct on all left cosets of $K$.  Given a generating set of $K$, there is at least one generator of the form $(k_1, k_2, 1)$. Pick $k_2$ to be the coset representative $u$ of $H$.  Form a generating set $S$ of $H$ as follows:  $S$ is initially empty.  For each generator of $K$, if it is of the form $(k_1, k_2, 0)$, then add $k_1$ and $uk_2u^{-1}$ to $S$; if it is of the form $(k_1, k_2, 1)$, then add $uk_1$ and $k_2u^{-1}$ to $S$. 
\end{proof}

\begin{corollary}
{\HC} has polynomial quantum query complexity.
\end{corollary}

It was mentioned in Friedl et al.\ \cite{friedl03:_hidden} that {\HC} in general is of exponential (classical) query complexity.

Using a similar approach, we show {\GHSh} essentially addresses {\HS} over a different family of groups.  We directly embed an instance of $(n,G)$-{\ghsh} into an instance of {\HS} over the group $G \wr \ints_n$.  When $n=2$, we get a polynomial-time reduction from {\HSh} over $G$ to {\HS} over $G \wr \ints_2$ (Corollary~\ref{cor:HShtoHS}).  This reduction was claimed independently (without proof) by Childs and Wocjan \cite{childs05}.

\begin{proposition}\label{GHS}
For $n\ge 2$ and $G$ a group, $(n,G)$-{\ghsh} reduces to {\HS} over $G\wr\ints_n$ in time polynomial in $n + s$, where $s$ is the size of the representation of an element of $G$.  Further, each instance of $(n,G)$-{\ghsh} can be recovered in polynomial time from its image under the reduction.
\end{proposition}

\begin{proof}
The input for {\GHSh} is a group $G$ and $n$ injective functions $f_1,
f_2, \ldots, f_n$ defined on $G$ such that for all $g \in G$,
$f_1(g)=f_2(ug), \ldots, f_{n-1}(g)=f_n(ug)$.  Consider the group $G
\wr \ints_n$.  Define a function $f$ such that for any element in
$(g_1,\ldots,g_n, \tau) \in G \wr \ints_n$, $f((g_1,\ldots,g_n,
\tau))=(f_{\tau(1)}(g_1),\ldots, f_{\tau(n)}(g_n))$.  The function
values of $f$ will be constant and distinct for right cosets of the
$n$-element cyclic subgroup generated by $(u, u, \ldots, u, u^{1-n},
1)$.

Given the $f$ defined in the last paragraph, it is trivial to recover
the original functions $f_1,\ldots,f_n$ by noting that $f_i(g)$ is the
$i$'th component of $f((g,\ldots,g,0))$.
\end{proof}

\begin{corollary}\label{cor:HShtoHS}
{\HSh} reduces to {\HS} in polynomial time (for arbitrary groups).
\end{corollary}

\begin{proof}
This is the $n=2$ case of Proposition~\ref{GHS}.
\end{proof}

Van Dam, Hallgren, and Ip \cite{dam03:_quant} introduced the Shifted
Legendre Symbol problem as a natural instance of {\HSh}.  They claimed
that assuming a conjecture this problem can also be reduced to an
instance of {\HS} over dihedral groups.  By
Corollary~\ref{cor:HShtoHS}, this problem can be reduced to {\HS} over
wreath product groups without any conjecture.

The case where $n>2$ in Proposition~\ref{GHS} may be more interesting
from a structural point of view then a complexity theoretic one.  We
already know \cite{childs05:_quant} that $(n,G)$-{\ghsh} for $n>2$
trivially reduces to $(2,G)$-{\ghsh}, simply by ignoring the
information provided by the functions $f_3,\ldots,f_n$.  One then gets
a polynomial-time reduction from $(n,G)$-{\ghsh} to {\HS} over
$G\wr\ints_2$.  Therefore, the reduction in Proposition~\ref{GHS} of
$(n,G)$-{\ghsh} to {\HS} over $G\wr\ints_n$ only tells us something
complexitywise if the instances of {\HS} over $G\wr\ints_n$ produced
by the reduction turn out to be \emph{easier} than those of {\HS} over
$G\wr\ints_2$.  This is conceivable, albeit unlikely.  Nonetheless,
the fact that $(n,G)$-{\ghsh} embeds into {\HS} over $G\wr\ints_n$ in
a natural way is interesting in itself, and may suggest other
reductions in a similar vein.

We also note that, unfortunately, it does not seem as though Proposition~\ref{GHS} translates the results of \cite{childs05:_quant} into fast quantum algorithms for any new family of instances of {\HS} over wreath product groups of the form $\ints_N \wr \ints_M$, because their algorithm is efficient only if $M \ge N^{\epsilon}$ for fixed $\epsilon > 0$, and our reduction is efficient only if $M$ is polylogarithmic in $N$.

Next we show that {\OC} is not a more general problem than {\HS}
either, if we allow the function in {\HS} to be a quantum function.
We need this generalization since the definition of {\OC} involves
quantum functions, i.e., the ranges of the functions are sets of
orthogonal quantum states.  In {\HS}, the function is implicitly
considered by most researchers to be a classical function, mapping
group elements to a classical set.  For the purposes of quantum
computation, however, this generalization to quantum functions is
natural and does not affect any existing quantum algorithms for {\HS}.

\begin{proposition}
{\OC} is quantum polynomial-time equivalent to {\HS}. 
\end{proposition}

\begin{proof}
Let $G$ and two orthogonal quantum states $\ket{\phi_0}, \ket{\phi_1}
\in \Gamma$ be the inputs of {\OC}. Define the function $\map{f}{G \wr
  \ints_2}{\Gamma\otimes\Gamma}$ as follows:
\[ f(g_1, g_2, \tau) = \left\{ \begin{array}{cl}
    \ket{g_1\cdot \phi_0} \otimes \ket{g_2\cdot \phi_1} & \mbox{if $\tau = 0$,}\\
    \ket{g_2\cdot \phi_1} \otimes \ket{g_1\cdot \phi_0} & \mbox{if $\tau = 1$.}
\end{array} \right.  \]  
The values of the function $f$ are identical and orthogonal on each left
coset of the following subgroup $H$ of $G \wr \ints_2$: If there is no
$u\in G$ such that $\ket{u \cdot 
  \phi_1} =\ket{\phi_0}$, then $H = G_{\ket{\phi_0}} \times G_{\ket{\phi_1}} \times \{0\}$. If there is such a $u$, then $H = (G_{\ket{\phi_0}}\times G_{\ket{\phi_1}} \times \{0\}) \cup
(G_{\ket{\phi_1}}u^{-1} \times uG_{\ket{\phi_1}}\times \{1\})$. 
For $i,j\in\{0,1\}$, let $g_i\in G$ be the $i$'th
coset representative of $G_{\ket{\phi_0}}$ (i.e., $\ket{g_i \cdot
    \phi_0}=\ket{\phi_i}$), and let $g_j\in G$ be the $j$'th coset
  representative of $G_{\ket{\phi_1}}$ (i.e., $\ket{g_j \cdot \phi_1}=\ket{\phi_j}$).
Then elements of the left coset of $H$ represented by
  $(g_i, g_j, 0)$ will all map to the same value
  $\ket{\phi_i}\otimes \ket{\phi_j}$ via $f$.
\end{proof}

\section{Decision versus Search} \label{search}

For any $\NP$-complete problem, its decision version and search
version are polynomial-time equivalent.  Another problem having this
property is {\GI} \cite{mathon79}. 

\subsection{{\HS} over permutation groups}
We adapt techniques in Arvind and Tor\'{a}n \cite{arvind01:_nc} to
show that over permutation groups, {\HS} also has this property.  

\begin{lemma} \label{multi}
Given (generating sets for) a group $G \leq S_n$, a function $\map{f}{G}{S}$ that hides a subgroup
$H\le G$, and a sequence of subgroups $G_1, \ldots, G_k \le S_n$, an instance of {\HS} can be
constructed to hide the group $D = \{(g,g,\ldots,g) \mid g \in H\cap G_1 \cap \cdots \cap G_k \}$ inside $G\times G_1\times\cdots\times G_k$.
\end{lemma}

\begin{proof}
Define a function $f'$ over the direct product group $G \times G_1 \times \cdots \times
G_k$ so that for any element $(g, g_1, \ldots, g_k)$, \ $f'(g, g_1,
\ldots, g_k) = (f(g), gg_1^{-1}, \ldots, gg_k^{-1})$. The values of
$f'$ are constant and distinct over left cosets of $D$.
\end{proof}

In the following, we will use the tuple $\tuple{G, f}$ to represent a standard
{\HS} input instance, and $\tuple{G, f, G_1, \ldots, G_k}$
to represent a {\HS} input instance constructed as in Lemma~\ref{multi}.

We define a natural isomorphism that identifies $S_n \wr \ints_2$ with a subgroup of $S_{\Gamma}$, where
$\Gamma=\{(i,j)\mid i \in \{1,\ldots,n\}, \;
j\in \{1,2\}\}$.  This isomorphism can be viewed as a group action, where the
group element $(g_1, g_2, \tau)$ maps $(i,j)$ 
to $(g_j(i), \tau (j))$.
Note that this isomorphism can be efficiently computed in both directions.

\begin{theorem}\label{decisionHS}
Over permutation groups, {\HS}$_S$ is truth-table reducible to
{\HS}$_D$ in polynomial time.
\end{theorem}

\begin{proof}
Suppose $f$ hides a nontrivial subgroup $H$ of $G$, first we compute
a strong generating set for $G$, corresponding to the chain $\{id\}=G^{(n)}\leq G^{(n-1)}\leq \cdots
\leq G^{(1)} \leq G^{(0)} = G$.  Define $f'$ over $G \wr
\ints_2$ such that $f'$ maps $(g_1, g_2, \tau)$ to $(f(g_1), f(g_2))$
if $\tau$ is 0, 
and $(f(g_2), f(g_1))$ otherwise. It is easy to check that for the
group $G^{(i)}\wr \ints_2$, $f'|_{G^{(i)}\wr\ints_2}$ hides
the subgroup $H^{(i)}\wr \ints_2$.

Query the {\HS}$_D$ oracle with inputs 
\[ \bigtuple{G^{(i)} \wr\ints_2,
f'|_{G^{(i)} \wr\ints_2}, (S_{\Gamma})_{\{(i,1), (j,2)\}},
(S_{\Gamma})_{\{(i,2), (j',1)\}}, (S_{\Gamma})_{\{(k,1), (\ell,2)\}}} \] 
for all $1\leq i \leq n$, all $j, j'\in\{i+1,\ldots,n\}$, and all $k,\ell \in \{i,\ldots, n\}$.

\begin{claim}
Let $i$ be such that $H^{(i)}=\{id\}$ and $H^{(i-1)} \neq \{id\}$. 
For all $i < j,j' \le n$ and all $i\le k,l \le n$, there is a (necessarily unique) permutation $h\in H^{(i-1)}$ such that
$h(i)=j$, $h(j')=i$ and $h(k)=\ell$ if and only if the query 
\[ \bigtuple{G^{(i-1)} \wr\ints_2,
f'|_{G^{(i-1)} \wr\ints_2}, (S_{\Gamma})_{\{(i,1), (j,2)\}},
(S_{\Gamma})_{\{(i,2), (j',1)\}}, (S_{\Gamma})_{\{(k,1), (\ell,2)\}}} \]
to the {\HS}$_D$ oracle answers ``nontrivial.''
\end{claim}

\begin{proofof}{Claim}
For any $j>i$, there is at most one permutation in $H^{(i-1)}$ that
maps $i$ to $j$.  To see this, suppose there are two distinct $h, h' \in
H^{(i-1)}$ both of which map $i$ to $j$.  Then $h'h^{-1} \in
H^{(i)}$ is a nontrivial permutation, contradicting the assumption
$H^{(i)}=\{id\}$. Let $h \in H^{(i-1)}$ be a permutation such that
$h (i)=j$, $h(j')=i$, and $h(k)=\ell$.  Then $(h,h^{-1},1)$ is a nontrivial
element in the group $H^{(i-1)}\wr \ints_2 \cap (S_{\Gamma})_{\{(i,1), (j,2)\}}
\cap (S_{\Gamma})_{\{(i,2), (j',1)\}} \cap (S_{\Gamma})_{\{(k,1), (\ell,2)\}}$, and 
thus the oracle answers ``nontrivial.''

Conversely, if the oracle answers ``nontrivial,'' then the nontrivial
element must be of the form $(h, h', 1)$ where $h,h'\in H^{(i-1)}$, 
since the other form $(h, h', 0)$ will imply that $h$ and $h'$ both fix $i$
and thus are in $H^{(i)} = \{id\}$. Therefore, $h$ will be a nontrivial element of 
$H^{(i-1)}$ with $h(i)=j$, $h(j')=i$, and $h(k)=\ell$.  This proves the Claim.
\end{proofof} 

Find the largest $i$ such that the query answers ``nontrivial'' for some $j,j'> i$ and some $k, \ell \ge i$.
Clearly this is the smallest $i$ such that $H^{(i)}=\{id\}$.
A nontrivial permutation in $H^{(i-1)}$ can be constructed by looking at the
query results that involve $G^{(i-1)} \wr \ints_2$.
\end{proof}

\begin{corollary}
Over permutation groups, {\HS}$_D$ and {\HS}$_S$ are polynomial-time equivalent.
\end{corollary}  

Next we show that the search version of {\HSh}, as a special case of {\HS}, also reduces to the corresponding decision problem.

\begin{definition}\label{def:HSh}\rm
Given a generating set for a group $G$ and two injective functions $f_1, f_2$ defined on $G$, the problem {\HSh}$_D$ is to determine whether there is a shift $u\in G$ such that $f_1(g) = f_2(gu)$ for all $g\in G$.
\end{definition}

\begin{theorem}
Over permutation groups, {\HSh}$_D$ and {\HSh}$_S$ are polynomial-time equivalent.
\end{theorem}

\begin{proof}
We show that if there is a translation $u$ for the two injective functions
defined on $G$, we can find $u$ with the help of an
oracle that solves {\HSh}$_D$.
First compute the strong generator set
$\cup_{i=1}^{n} C_i$ of $G$ using the procedure in \cite{furst80:_polyn}.
Note that $\cup_{i=k}^{n} C_i$ generates $G^{(k-1)}$ for $1\leq k \leq n$.
We will proceed in steps along the stabilizer subgroup chain
$G=G^{(0)}\geq G^{(1)} \geq \cdots \geq G^{(n)}=\{id\}$.

\begin{claim}
With the help of the {\HSh}$_D$ oracle, finding the translation $u_i$ for
input $(G^{(i)}, f_1, f_2)$ reduces to finding another translation
$u_{i+1}$ for input $(G^{(i+1)}, f_1', f_2')$. In particular, we have
$u_i=u_{i+1}\sigma_i$. 
\end{claim}
\begin{proofof}{Claim}
Ask the oracle whether there is a translation for input
$(G^{(i+1)}, f_1|_{G^{(i+1)}}, f_2|_{G^{(i+1)}})$. If the answer is
yes, then we know $u_i\in G^{(i+1)}$ and therefore set $\sigma_i=id$
and $u_{i}=u_{i+1}\sigma_i$. 

If the answer is
no, then we know that $u$ is in some right coset of $G^{(i+1)}$ in
$G^{(i)}$. 
For every $\tau \in C_{i+1}$, define a function $f_{\tau}$
such that $f_{\tau}(x)=f_2(x\tau)$ for all $x\in G^{(i+1)}$. Ask the
oracle whether there is a translation for input $(G^{(i+1)},
f_1|_{G^{(i+1)}}, f_{\tau})$. The oracle will answer yes if and only
if $u$ and $\tau$ are in the same right coset of $G^{(i+1)}$ in
$G^{(i)}$, since
\begin{eqnarray*}
& &  \mbox{$u$ and $\tau$ are in the same right coset of
  $G^{(i+1)}$ in $G^{(i)}$}  \\
& \Longleftrightarrow & u=u'\tau \mbox{ for some $\tau'\in
  G^{(i+1)}$}\\
& \Longleftrightarrow & f_1(x)=f_2(xu)=f_2(xu'\tau)=f_{\tau}(xu') \mbox{ for all $x\in G^{(i)}$}\\
& \Longleftrightarrow & \mbox{$u'$ is the translation for $(G^{(i+1)},
  f_1|_{G^{(i+1)}}, f_{\tau})$.}
\end{eqnarray*}
Then we set $\sigma_i=\tau$. 
\end{proofof}

We apply the above procedure $n-1$ times until we reach the trivial
subgroup $G^{(n)}$. The translation $u$ will be equal to
$\sigma_n\sigma_{n-1}\cdots\sigma_1$. Since the size of each $C_i$ is
at most $n-i$, the total reduction is in classical polynomial time.  
\end{proof}

\subsection{{\HS} over dihedral groups}

For {\HS} over dihedral groups $D_n$, we can efficiently reduce search to decision when $n$ has small prime factors.
For a fixed integer $B$, we say an
integer $n$ is $B$-smooth if all
the prime factors of $n$ are less than or equal to $B$. For such an $n$,
the prime factorization can be obtained in time polynomial in $B + \log n$. 
Without loss of generality, we assume that the hidden
subgroup is an order-two subgroup of $D_n$ \cite{ettinger00}.

\begin{theorem}
Let $n$ be a $B$-smooth number, {\HS} over the dihedral group $D_n$
reduces to {\HS}$_D$ over dihedral groups in time polynomial in $B + \log n$.
\end{theorem}

\begin{proof}
Without loss of generality, we assume the generator set for $D_n$ is
$\{r, \sigma\}$, where the order of $r$ and $\sigma$ are $n$ and 2,
respectively. Let $p_1^{e_1}p_2^{e_2}\cdots p_k^{e_k}$ be the prime
factorization of $n$. Since $n$ is $B$-smooth, $p_i \leq B$ for all
$1\leq i \leq k$. Let the hidden subgroup $H$ be $\{id, r^a\sigma\}$
for some $a < n$. 

First we find $a \bmod p_1^{e_1}$ as follows. Query the {\HS}$_D$
oracle with input groups
(we will always use the original input function $f$) 
$\tuple{r^{p_1}, \sigma}, \tuple{r^{p_1}, r\sigma}, \ldots, \tuple{r^{p_1},
r^{p_1-1}\sigma}$. It is not hard to see that the {\HS}$_D$ oracle
will answer ``nontrivial'' only for the input group $\tuple{r^{p_1},
r^{m_1}\sigma}$ where $m_1 = a \bmod p_1$. The next set of input groups
to the {\HS}$_D$ oracle are $\tuple{r^{p_1^2}, r^{m_1}\sigma},  \tuple{r^{p_1^2},
r^{p_1+m_1}\sigma}, \ldots, \tuple{r^{p_1^2}, r^{(p_1-1)p_1+m_1}\sigma}$. From the
oracle answers we obtain $m_2 = a \bmod p_1^2$. Repeat the above
procedure until we find $a \bmod p_1^{e_1}$.  

Similarly, we can find $a \bmod p_2^{e_2}, \ldots, a \bmod p_k^{e_k}$. 
A simple usage of the Chinese Remainder Theorem will then recover $a$.
The total number of queries is $e_1p_1+e_2p_2+\cdots +e_kp_k$, which
is polynomial in $\log n + B$.
\end{proof}

\section{Nonadaptive Checkers}\label{programchecker}

An important concept closely related to self-reducibility is
that of a \emph{program checker}, which was first introduced by Blum and
Kannan \cite{blum95:_desig}. They gave program checkers for some
group-theoretic problems and selected problems in $\P$. They also
characterized the class of problems having polynomial-time checkers.
Arvind and Tor\'{a}n \cite{arvind01:_nc} presented a nonadaptive $\NC$
checker for {\Gint} 
over permutation groups.
In this section we show that {\HS}$_D$ and {\HS} over permutation groups 
have nonadaptive checkers.

For the sake of clarity, we give the checker for {\HS}$_D$ first.
Let $P$ be a program that solves {\HS}$_D$ over permutation groups. The input
for $P$ is a permutation group $G$ given by its generating set and a function
$f$ that is defined over $G$ and hides a subgroup $H$ of $G$. If $P$ is a correct
program, then $P(G,f)$ outputs TRIVIAL if $H$ is the trivial subgroup of $G$, and 
NONTRIVIAL otherwise. The checker $C^P(G,f,0^k)$ checks the program $P$ on the input
$G$ and $f$ as follows:

\vspace{.25cm}
\small
{\bf Begin}

Compute $P(G,f)$.

{\bf if} $P(G,f)=$ NONTRIVIAL, {\bf then}

\hspace*{.5cm}Use Theorem~\ref{decisionHS} and $P$ (as if it were bug-free) to 
search for a nontrivial element $h$ of $H$.

\hspace*{.5cm}{\bf if} $f(h)=f(id)$, {\bf then}

\hspace*{1cm}{\bf return} CORRECT

\hspace*{.5cm}{\bf else}

\hspace*{1cm}{\bf return} BUGGY

{\bf if} $P(G,f)=$ TRIVIAL, {\bf then}

\hspace*{.5cm}{\bf Do} $k$ times (in parallel):

\hspace*{1cm}generate a random permutation $u\in G$.

\hspace*{1cm}define $f'$ over $G$ such that $f(g)=f'(gu)$ for all
$g\in G$, use $(G, f, f')$ to be an input instance of {\HSh}

\hspace*{1cm}use Theorem~\ref{HC} to convert $(G,f,f')$ to an input
instance $(G\wr \ints_2, f'')$ of {\HS}

\hspace*{1cm}use Theorem~\ref{decisionHS} and $P$ to search for a nontrivial element $h$ 
of the subgroup of $G\wr \ints_2$ that $f''$ hides.

\hspace*{1cm}{\bf if} $h \neq (u^{-1}, u, 1)$, {\bf then} {\bf return} BUGGY 

\hspace*{.5cm}{\bf End-do}

\hspace*{.5cm}{\bf return} CORRECT

{\bf End}
\normalsize
\vspace{.25cm}

\begin{theorem}\label{checker}
If $P$ is a correct program for {\HS}$_D$, then $C^P(G,f,0^k)$ always outputs CORRECT\@.
If $P(G,f)$ is incorrect, then $\Pr[\mbox{$C^P(G,f,0^k)$ outputs CORRECT}]\leq 2^{-k}$.  Moreover, 
$C^P(G,f,0^k)$ runs in polynomial time and queries $P$ nonadaptively. 
\end{theorem}
\begin{proof}
If $P$ is a correct program and $P(G,f)$ outputs NONTRIVIAL, then 
$C^P((G,f,0^k)$ will find a nontrivial element of $H$ and outputs CORRECT. If $P$ 
is a correct program and $P(G,f)$ outputs TRIVIAL, then the function $f'$ constructed
by $C^P(G,f,0^k)$ will hide the two-element subgroup $\{(id,id,0),(u, u^{-1},1)\}$. 
Therefore, $C^P(G,f,0^k)$ will always recover the random permutation $u$ correctly, and 
output CORRECT. 

On the other hand, if $P(G,f)$ outputs NONTRIVIAL while $H$ is actually trivial, 
then $C^P(G,f,0^k)$
will fail to find a nontrivial element of $H$ and thus output BUGGY. 
If $P(G,f)$ outputs TRIVIAL while $H$ is actually nontrivial, then the function $f''$ 
constructed
by $C^P(G,f,0^k)$ will hide the subgroup $(H\,\times\, u^{-1}Hu\,\times\, \{0\})\cup (u^{-1}H\,\times\, Hu\,times\, \{1\})$.
$P$ correctly distinguishes $u$ and other
elements in the coset $Hu$ only by chance. Since the order of $H$ is at least 2, 
the probability that $C^P(G,f,0^k)$ outputs CORRECT is at most $2^{-k}$.

Clearly, $C^P(G,f,0^k)$ runs in polynomial time. The nonadaptiveness follows from 
Theorem~\ref{decisionHS}.
\end{proof}

Similarly, we can construct a nonadaptive  
checker $C^P(G,f,0^k)$ for a program $P(G,f)$ that solves {\HS} over permutation
groups. The checker makes $k$ nonadaptive 
queries.

\vspace{.25cm}
\small
{\bf Begin}

Run $P(G,f)$, which outputs a generating sets $S$.

Verify that elements of $S$ are indeed in $H$.

{\bf Do} $k$ times (in parallel):

\hspace*{.5cm}generate a random element $u\in G$.

\hspace*{.5cm}define $f'$ over $G$ such that $f(g)=f'(gu)$ for all
$g\in G$, use $(G, f, f')$ to be an input instance of {\HC} 

\hspace*{.5cm}use Theorem~\ref{HC} to convert $(G,f,f')$ to an input
instance $(G\wr \ints_2, f'')$ of {\HS}
 
\hspace*{.5cm}$P(G\wr\ints_2, f'')$ will output a set $S'$ of
generators and a coset representative $u'$

\hspace*{.5cm}{\bf if} $S$ and $S'$ don't generate the same group {\bf
  or} $u$ and $u'$ are not in the same coset of $S$, {\bf then}

\hspace*{.5cm}{\bf return} BUGGY
 
{\bf End-do}

{\bf return} CORRECT

{\bf End}
\normalsize
\vspace{.25cm}

The proof of correctness for the above checker is very similar to the
proof of Theorem~\ref{checker}.

\section{Further Research}

Each of the problems we have looked at in this paper can vary widely
in complexity, depending on the type underlying group.  So it is, for
instance, with {\HS}, which yields to quantum computation in the
abelian case but remains apparently hard in all but a few nonabelian
cases.  The reductions of these problems to {\HS} given in this paper
all involve taking wreath products, which generally increases both the
size and the ``difficulty'' of the group considerably.  (For example,
$G\wr H$ is never abelian unless one of the groups is abelian and the
other is trivial, whence $G\wr H \cong G$ or $G\wr H \cong H$.)  It is
useful in general to find reductions between these problems that map
input groups to output groups that are of similar difficulty, e.g.,
abelian $\mapsto$ abelian, solvable $\mapsto$ solvable, etc.  This
would provide a finer classification of the complexities of these
problems.

The embedding aspect of the reduction in Proposition~\ref{GHS}
suggests a stronger question: given \emph{any} function $f$ on $G\wr
\ints_n$ that hides some subgroup generated by
$(u,\ldots,u,u^{1-n},1)$ for some $u$ (where the function is not
necessarily the one constructed by the reduction), can one efficiently
recover an instance of $(n,G)$-{\ghsh} that maps via the reduction to
an instance of {\HS} over $G\wr\ints_n$ with the same hidden subgroup?
A yes answer would show that {\GHSh} is truly a special case of {\HS},
and as a corollary would show that these instances of {\HS} over
$G\wr\ints_n$ for small $n$ (polynomial in the size of elements of
$G$) reduces to {\HS} over $G \wr \ints_2$.

\section{Acknowledgments}

We thank Andrew Childs and Wim van Dam for valuable comments on a
preliminary version of this paper.

\end{document}